\documentclass[preprintnumbers,amsmath,amssymb]{revtex4}

\newcommand{\be}{\begin{equation}}
\newcommand{\ee}{\end{equation}}
\newcommand{\ben}{\begin{eqnarray}}
\newcommand{\een}{\end{eqnarray}}

\newcommand{\nd}{\noindent}
\usepackage{amsmath}
\usepackage{graphicx}% Include figure files
\usepackage{dcolumn}% Align table columns on decimal point
\usepackage{bm}% bold math
\usepackage[usenames]{color}
\usepackage{multirow}

\begin{document}

\preprint{draft version}
\title{New features of quantum discord uncovered by q-entropies}

\author{A.P. Majtey$^{1}$\footnote{anamajtey@ugr.es},
 A.R. Plastino$^{1,\,2}$\footnote{arplastino@ugr.es}, and A. Plastino$^{1,\,3}$\footnote{angeloplastino@gmail.com}}
\affiliation{ $^1$Instituto Carlos I de F\'{\i}sica Te\'orica y
Computacional and Departamento de F\'{\i}sica At\'omica Molecular
y Nuclear, Universidad de Granada, Granada 18071, Spain \\\\
$^2$National University La Plata UNLP-CREG, C.C. 727, 1900
La Plata, Argentina\\\ $^3$National University La Plata
UNLP-IFLP-Conicet, C.C. 727, 1900 La Plata, Argentina}

\date{\today}
\begin{abstract}

\nd  The notion of quantum discord introduced by Ollivier and Zurek
 [Phys. Rev. Lett {\bf 88}, 017901 (2001)] (see also
Henderson and Vedral [J. Phys. A {\bf 34}, 6899 (2001)])
has attracted  increasing attention, in recent years, as  an entropic
 quantifier of non-classical features pertaining to the correlations exhibited by
 bipartite quantum systems. Here we generalize the notion so as to
 encompass power-law q-entropies (that reduce to the standard Shannon entropy
 in the limit $q \to 1$) and study the concomitant consequences. The
 ensuing, new discord-like measures we advance describe aspects of non-classicality
 that are different from those associated with the standard quantum discord.
 A particular manifestation of this difference concerns a feature related to
 order. Let $D_1$ stand for the standard, Shannon-based discord measure and
 $D_q$ for the $q \ne 1$ one.  If two quantum states $A,\,B$ are such that
 $D_1(A) > D_1(B)$, this order-relation does not remain invariant under a
 change from $D_1$ to $D_q$.

\end{abstract}

\pacs{03.67.-a; 03.67.Mn; 03.65.-w}

\maketitle

%\newpage

\section{Introduction}

\nd The degree of understanding of quantum correlations (QC) underlies our current picture of Nature
\cite{AFOV08,BEN}. It has been recently found that there exist important manifestations of the
quantumness of correlations in composite systems that are different from those
of entanglement-origin (EO) and that may be relevant in quantum information technologies
\cite{DVB10,OZ02,HV01,CABMPW10}.
The quantifier of these non-EO correlations is called the quantum discord (QD) $D_1$
 and is based, for a bipartite system, on  Shannon's mutual
information. We are thus speaking of an information-theoretic (IT) tool.
For pure states QD does not add any QCs, but that is not the case for mixed states.
 The $D_1-$concept, advanced in the pioneer paper by Ollivier and Zurek \cite{OZ02},  quantifies:
i) the minimum change in the state of the system and ii)  the
information on one of its parts induced by a measurement of the
other one. If the state has only classical correlations, $D_1$
vanishes, which implies that the quantum discord concept somehow
quantifies the ``correlational-quantumness''. It has been
evaluated for several families of states both in its original form
and in variously altered versions and generalizations. A
particularly compelling  instance expresses the QD-notion in terms
of conditional density operators \cite{LLZ07}. Interesting
operational QD-interpretations have also been advanced
\cite{CABMPW10}.
 Evaluating QD requires a rather involved  optimization procedure, analytical expressions being known in just a few instances \cite{Luo08,Luo08b,LL08,Vedral03,MCSV09,OHHH02,ARA10}. $D_1$ is built up as an entropic difference,
the difference between a quantum entropic measure  and its classical counterpart,
which is derived from local measurements on one or both of the participant subsystems. Its amount is a new feature
$D_1(A)$ of the quantum state $A$, which in turn induces a $D_1-$amount ``ordering'' for states of the form $D_1(A)> D_1(B)$, for instance.

\vskip 3mm \nd Now, IT-tools come in many distinct varieties. Given the
immense body of literature that has been generated in the past two decades
concerning physically motivated statistical formalisms based on generalizations
of Shannon's information measure (see \cite{Tsallislibro,GT2004,B09,N11}  and references
therein), it seems both natural and necessary to tackle the QD
issues from this generalized angle, in the hope of gaining interesting insights,
and, in particular, so as to establish the invariance or not of
the discord-induced order under a change of the prevailing statistics,
from Shannon's to its many rivals.

\vskip 3mm \nd To show that this is indeed a fruitful endeavor is
the  aim of the present paper, in which a generalization of the
quantum discord {\it concept}, in the context of generalized
statistics, will be advanced and the ``ordering-question''
answered. In section 2 we introduce our conceptual QD-extension,
discussing its main properties in Sect. 3. Next  we present some
results for general bipartite states, focusing attention on
analytical results.  We also perform numerical simulations for
random bipartite states (Sect. 4). Finally, some conclusions are
drawn in section 5.

\section{Retracing the Ollivier and Zurek's path \`a la Tsallis for getting a quantum $\text{q}$-discord}

\nd Tsallis' power-law $q$-entropy was introduced in \cite{Tsallis88}
as an extension of the Shannon entropy as follows \cite{Tsallislibro}

\begin{equation}
 H_q(X)=-\sum_x p(x)^q\ln_qp(x),
\end{equation}

\noindent where the $q$-logarithm is defined by $\ln_q(x)\equiv\frac{x^{1-q}-1}{1-q}$, $p(x)\equiv p(X=x)$ is the
probability distribution of the pertinent random variable $X$, and $q$
is any nonnegative real number. Tsallis's entropy converges to  Shannon's
in the limit $q\to 1$. $H_q$ plays a fundamental  role  in recent developments
of statistical mechanics \cite{Tsallislibro,GT2004,B09,N11,ASMNC10}.
The generalization has indeed received a lot of attention in the last years,
with about 2000 papers containing interesting results and
useful applications, many of them in the complex systems'
area \cite{Tsallislibro,GT2004,B09,N11,ASMNC10} but also in
connection with a variegated family of quantum mechanical settings (see, for example
\cite{NRT11,L11,AALE2011,SAA10,H10,HPP2009,N09,CT08,ZP06,BCPP02,MML02,TBD98}).
In what follows we retrace the developments of Ref. \cite{OZ02} in a Tsallis, $q-$context.
Thus, just by setting $q=1$ we recover the Ollivier-Zurek quantities. We begin then
with the mutual information, defined as

\begin{equation}
 I_q(X:Y)=H_q(X)+H_q(Y)-H_q(X,Y),
\end{equation}
and the following chain rule holds \cite{Furuichi06}:

\begin{equation}
 H_q(X,Y)=H_q(Y)+H_q(X|Y),
\end{equation}

\noindent where the conditional entropy reads
\begin{equation}
H_q(X|Y)=\sum_y p(y)^qH_q(X|y).
\end{equation}
The chain rule gives the relation between a conditional entropy and a joint entropy.
Using this relation we can define another, classically equivalent, expression for the mutual information

\begin{equation}
 J_q(X:Y)=H_q(X)-H_q(X|Y).
\end{equation}
 The $I\,-\,J$ difference is of the essence for Ollivier-Zureck goals, after expressing the two quantifiers
  in quantal fashion.
   Let us then do the same with   $I_q$ and $J_q$.  $I_q$ can be easily generalized
defining  appropriate density matrices for the quantum systems,
$\rho_A$, $\rho_B$, and $\rho_{A,B}$, and applying then the
$q$-generalization of  von Neumann's entropy $S_q(\rho)=-{\text
Tr} (\rho^q\ln_q\rho)$. One has

\begin{equation}\label{Imutua}
 I_q(A:B)=I_q(\rho_{A,B})=S_q(\rho_A)+S_q(\rho_B)-S_q(\rho_{A,B}).
\end{equation}
To generalize the  $J_q-$expression, following \cite{OZ02}, we
focus on a perfect measurements of $\rho_B$ defined by a set of
projectors $\{\Pi_j^{(B)}\}$ such that $\sum_j\Pi_j^{(B)}=1$.
Accordingly,

\begin{equation}\label{Jmutua}
 J_q(\rho_{A,B})_{\{\Pi_j^{(B)}\}}=S_q(\rho_A)-S_q(\rho_A|\{\Pi_j^{(B)}\}),
\end{equation}
\noindent where

\begin{equation}
 S_q(\rho_A|\{\Pi_j^{(B)}\})=\sum_j p_j^qS_q(\rho_{A|\Pi_j^{(B)}}),
\end{equation}
\noindent with the state of $A$ given, once the measurement is performed, by

\begin{equation}
 \rho_{A|\Pi_j^{(B)}}=\Pi_j^{(B)}\rho_{A,B}\Pi_j^{(B)}/{\text Tr}_{A,B}\Pi_j^{(B)}\rho_{A,B},
\end{equation}
\noindent and $p_j={\text Tr}_{A,B}\Pi_j^{(B)}\rho_{A,B}$.

\nd The two classical expressions for the standard mutual
information we have presented above are identical, but this is not
necessarily so in the quantum case Actually, the quantum discord
is defined as the {\it minimum} possible difference between the
two possibilities, given by an {\it optimum} set of
$\{\Pi_j^{(B)}\}$ \cite{OZ02}. Thus, we are to be  concerned here
with what happens to the expressions $I_q$ and $J_q$. Introduce
the quantity

\begin{equation}\label{optimization}
 C_q(\rho_{A,B}):=\sup_{\{\Pi_j^{(B)}\}} J_q(\rho_{A,B})_{\{\Pi_j^{(B)}\}}.
\end{equation}
\noindent We  define now our quantum q-discord as the difference

\begin{equation}
 \vartheta_q(\rho_{A,B})=I_q(\rho_{A,B})-C_q(\rho_{A,B}).
\end{equation}
This quantum q-discord is the minimum of the difference between Eq. (\ref{Imutua}) and Eq. (\ref{Jmutua}).
We normalize this measure via a trivial re-scaling in order to compare, in an adequate way, different quantities:

\begin{equation}
 D_q(\rho_{A,B})=\frac{q-1}{1-2^{1-q}}\vartheta(\rho_{A,B})
\end{equation}
For similar reasons, $\log$ denotes the logarithm of base 2. In
what follows $\rho_{A,B}\equiv\rho$. Note that $\lim_{q\to
1} D_q(\rho)=D_1(\rho)$.

\section{Properties of the generalized quantum discord}

\nd We see that $D_q\ge 0$ for $q\in(0,1)$, and this might be
related to the concavity of the $q$-conditional entropy
$S_q(\rho_{A,B})-S_q(\rho_B)$ with respect to $\rho_{A,B}$.
%(however this property has not been confirmed for Tsallis
%conditional entropy even for $q\in(0,1)$).
Indeed, $\vartheta_q$ becomes negative if $q$ grows from
$1\to\infty$, negativity increasing with $q$.  Quite convenient is
the particular case $q=2$, since it requires  only to compute the
traces of $\rho^2,\rho_B^2,\rho_k^2$, and matrix-diagonalization
is avoided, making  computations more efficient. Taking the limit
$q\to\infty$ of the normalized measure we obtain

\begin{equation}
\lim_{q\to\infty}D_q= \begin{cases} 0, & \mbox{mixed states, }  \\ 1 , & \mbox{pure states.}  \end{cases}
\end{equation}
To study the positivity of the q-discord we consider separately two cases: pure states and mixed states.

\subsubsection{Pure states}

\nd For pure states, that is $\rho=|\Psi\rangle\langle\Psi|$, the
q-discord takes the form $\vartheta(\rho)=S_q(\rho_A)$ and the
quantum q-discord coincides with the reduced (quantum) Tsallis
entropy. We can easily verify this fact by casting $|\Psi\rangle$
in its Schmidt decomposition form
$|\Psi\rangle=\sum_i\lambda_i|ii\rangle$. Thus, since the
q-entropy is positive for all $q$,

\begin{equation}
 \vartheta_q(|\Psi\rangle\langle\Psi|)\geq 0, \,\,\,\,\,\, \forall\, q.
\end{equation}

\subsubsection{Mixed states}

\nd In the case of mixed states our q-discord is positive only for
values of $q$ in $(0,1)$. In order to demonstrate the positivity of
the q-discord for mixed arbitrary states we  follow \cite{OZ02}
 and  consider  the proposition:
$S_q(\rho_A|\{\Pi_j^{(B)}\})=S_q(\rho_{A,B}^{(D)})-S_q(\rho_B^{(D)}),$
with $\rho_{A,B}^{(D)} =\sum_j p_j\rho_j$. Now, $\rho_{A,B}^{(D)}$
is block diagonal and as in the $q=1$ case, doing a block by block
analysis the proposition can be proved. Now we need to verify the
inequality:

\begin{equation}
S_q(\rho_A|\{\Pi_j^{(B)}\})\geq S_q(\rho_{A,B})-S_q(\rho_B).
\end{equation}
By recourse of the previous preposition we can establish  the
following relation for any measurement $\{\Pi_j^{(B)}\}$

\begin{equation}
S_q(\rho_A|\{\Pi_j^{(B)}\})=S_q(\rho_{A,B}^{(D)})-S_q(\rho_B^{(D)} ),
\end{equation}
and by the (conjectured) concavity (see below) of the conditional
entropy ($S_q(\rho)-S_q(\rho_B)$) with respect to $\rho$ for
$q\in(0,1)$ we are led to

\begin{equation}
S_q(\rho_{A,B}^{(D)} )-S_q(\rho_B^{(D)} )\geq S_q(\rho_{A,B})-S_q(\rho_B).
\end{equation}

\subsubsection{Random generation of states in an $N-$dimensional Hilbert space}

\nd  The set of  states in an $N$-dimensional Hilbert space can be regarded as a product-space of the
form \cite{ZHS98,PZK98},

\[ {\cal{H}}=\mathcal{P} \times \Delta,
\]
where $\mathcal{P}$ stands for the family of all complete sets of
ortho-normal projectors $\{\hat P_i\}_i^N,\,\,\sum_i \hat
P_i=\mathbb{I}$ ($\mathbb{I}$ the identity matrix), and $\Delta$
is the convex set of all real $N-$tuples of the form
$\{\lambda_1,\ldots,\lambda_N\};\,\,\lambda_i
\in\mathbb{R};\,\,\sum_i \lambda_i=1;\,\,0\le \lambda_i \le 1.$
Any state in  $\cal{H}$ takes the form $\rho=\sum_i \lambda_i \hat
P_i.$
\vskip 3mm
\nd In order to explore ${\cal{H}}$ we introduce an appropriate
measure $\mu $ on this space. Such a measure is required to
compute volumes within ${\cal{H}}$, as well as to determine what
is to be understood by a uniform distribution of states on ${\cal
H}$. An arbitrary state $\rho$
of our $N$-dimensional Hilbert space can always be expressed as a product
of the form

\begin{equation}
\label{udot} \rho=U D[\{\lambda_i\}] U^{\dagger}.
\end{equation}
\noindent Here $U$ is an $N\times N$ unitary matrix and
$D[\{\lambda_i\}]$ is an $N\times N$ diagonal matrix whose
diagonal elements are, precisely, our above defined  $\{\lambda_1,
\ldots,\lambda_N \}$. The group of unitary matrices $U(N)$ is
endowed with a unique, uniform measure, known as the Haar's
measure, $\nu$ \cite{PZK98}. On the other hand, the $N$-simplex
$\Delta$, consisting of all the real $N$-uples $\{\lambda_1,
\ldots, \lambda_N \}$ appearing in (\ref{udot}), is a subset of a
$(N-1)$-dimensional hyperplane of $\mathbb{R}^N$. Consequently,
the standard normalized Lebesgue measure ${\cal L}_{N-1}$ on
$\mathbb{R}^{N-1}$ provides a measure for $\Delta$. The
aforementioned measures on $U(N)$ and $\Delta$ lead to a measure
$\mu $ on the set ${\cal S}$ of all the states of our quantum
system \cite{ZHS98,PZK98},

\begin{equation} \label{memu}
 \mu = \nu\times {\cal L}_{N-1}.
\end{equation}
If one needs to randomly generate mixed states, this is to be done according to the measure
(\ref{memu}).

\subsubsection{Concavity of the conditional entropy in the interval $(0<q<1)$}

\nd Here we attempt a numerical verification of the concavity of
the conditional entropy for  $q\in(0,1)$, that is,

\begin{equation}\label{Deltaq}
 \Delta_q=S_q(\rho)-S_q(\rho_A) - \{t[S_q(\sigma)-S_q(\sigma_A)]+(1-t)[S_q(\xi)-S_q(\xi_A)]\}\geq
 0,
\end{equation}
where $\rho=t\sigma+(1-t)\xi$, $\rho_A$ is the reduced density
matrix corresponding to $\rho$, $\sigma_A$ ($\xi_A$) the reduced
density matrices of $\sigma$ ($\xi$) and, finally,  $0\leq t\leq
1$.

\begin{figure}[h]
\begin{center}
\vspace{0.5cm}
\includegraphics[scale=1,angle=0]{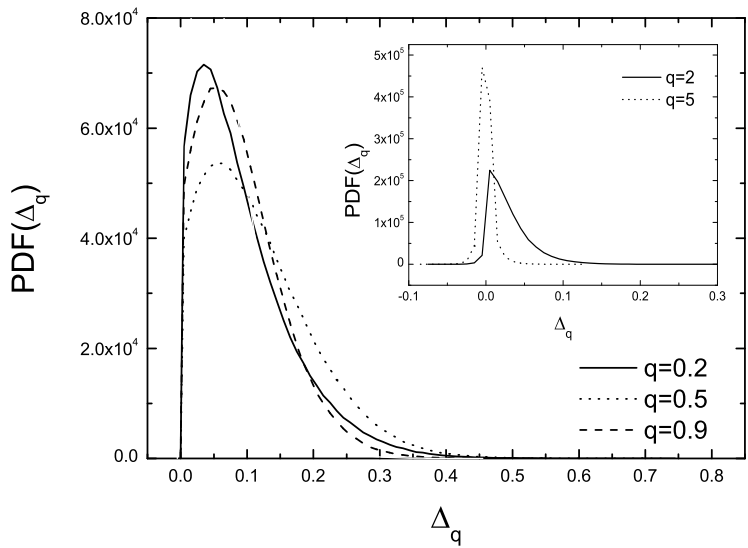}
\vspace{0.5cm} \caption{Probability distribution for $\Delta q$
for different values of $q<1$ for which the concavity is verified.
Inset: $PDF(\Delta_q)$ for different $q>1$. The curves are left-
shifted. All curves were constructed using of the order of $10^6$
(numerically generated) states. All depicted quantities are
dimensionless.\label{PDFs}}
\end{center}
\end{figure}

\nd The concavity of the standard conditional entropy ($q=1$) was
proved in \cite{Wehrl} by assuming the validity of a lemma by Lieb
\cite{Lieb}. The proof is rather difficult even in this case \cite{Wehrl}. As a first
 step we evaluate numerically the inequality (\ref{Deltaq}) by
generating random states in an $N$-dimensional Hilbert space.  In order to assess, for these randomly generated
states, how the concavity-requirement is satisfied, we evaluate
(\ref{Deltaq}) for a large enough number of simulated states
($\sigma$ and $\xi$). We set $N=4$ for the dimension of the
state-space in all simulations and we randomly generate $t\in
[0,1]$.

\nd We investigate the positivity of $\Delta_q$, upon which the
concavity of the conditional q-entropy is based, by constructing
the probability distributions for the values of $\Delta q$. The
corresponding distributions, for different values of $q$, are
depicted in Fig. \ref{PDFs}. In the inset we plot the probability
distribution of $\Delta_q$ for $q=2$ and $q=5$. The curves are
constructed using of the order of $10^6$ states. These simulations
provide us with strong evidence about the validity of the
conjecture advanced above on the concavity of the conditional
q-entropy and, consequently, on the positivity of the quantum
q-discord for $0 < q <1$.

\section{Relation between $q$-discord and orthodox discord}

\nd Let us now investigate  the relation between the q-discord and
its original counterpart for different sorts of states.

\nd We begin with  Bell diagonal states. These are  two-qubit states with
maximally-mixed reduced-density matrices and have  the form

\begin{equation}
 \rho_{A,B}=\rho=\frac{1}{4}\left(I+\sum_{j=1}^3c_j\sigma_j\otimes\sigma_j\right)
\end{equation}
where $c_j$ are real constants constrained by certain conditions
(in order to have a well defined density operator $\rho$) and
$\sigma_j$'s are the Pauli operators. Let
$\lambda_i=\lambda_i(c_j)\in[0,1]$, $(i=0,...,3)$ be the
eigenvalues of $\rho$

\begin{eqnarray}
 \lambda_0&=&\frac{1}{4}(1-c_1-c_2-c_3),\nonumber\\
 \lambda_1&=&\frac{1}{4}(1-c_1+c_2+c_3),\nonumber\\
 \lambda_2&=&\frac{1}{4}(1+c_1-c_2+c_3),\nonumber\\
 \lambda_3&=&\frac{1}{4}(1+c_1+c_2-c_3).
\end{eqnarray}
The marginal states of $\rho$ are $\rho_A={\mathbb I}/2$ and
$\rho_B={\mathbb I}/2$. Thus, the quantum q-mutual information of
$\rho$ is

\begin{equation}
 I_q(\rho)=-4\left(\frac{1}{2}\right)^q\ln_q\frac{1}{2}+\sum_i\lambda_i^q\ln_q\lambda_i
\end{equation}
and

\begin{equation}
 C_q(\rho)=2\left(\frac{1}{2}\right)^q\left[-\ln_q\frac{1}{2}+\left(\frac{1-c}{2}\right)^q
 \ln_q\frac{1-c}{2}+\left(\frac{1+c}{2}\right)^q\ln_q\frac{1+c}{2}\right],
\end{equation}
where $c:=\max\{|c_1|,|c_2|,|c_3|\}$. We find, for a general
(Bell-diagonal) two-qubit state,

\begin{equation} \label{ana1}
 \vartheta(\rho)=-2\left(\frac{1}{2}\right)^q\left[\ln_q\frac{1}{2}+\left(\frac{1-c}{2}\right)^q\ln_q\frac{1-c}{2}+\left(\frac{1+c}{2}\right)^q\ln_q\frac{1+c}{2}\right]+\sum_i\lambda_i^q\ln_q\lambda_i.
\end{equation}

\begin{figure}[h]
\begin{center}
\vspace{0.5cm}
\includegraphics[scale=0.75,angle=0]{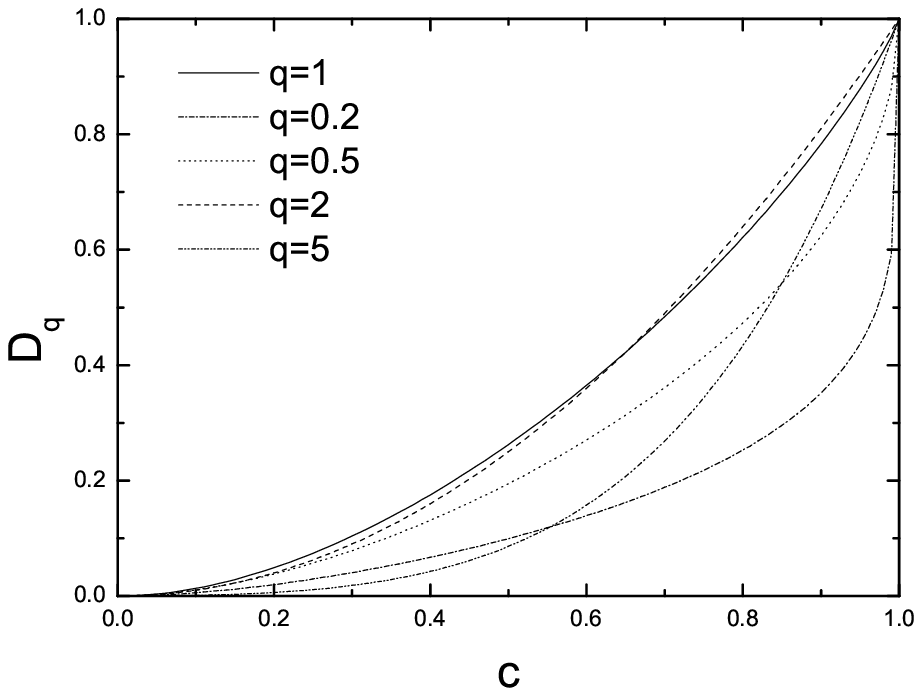}
\includegraphics[scale=0.75,angle=0]{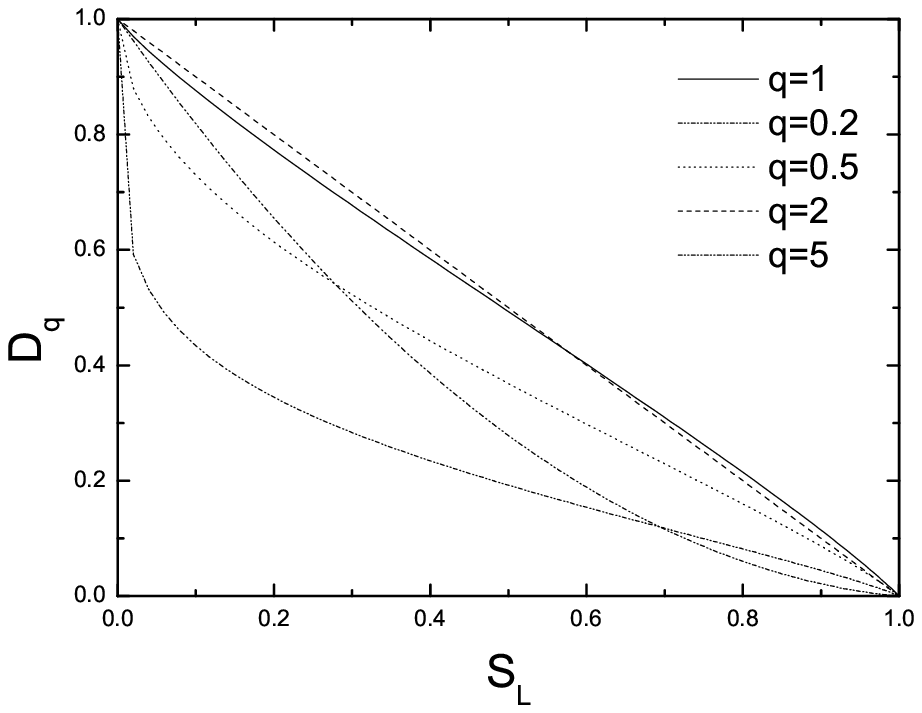}
\vspace{0.5cm} \caption{(a) $D_q$ for Werner state, as a function
of $c$, for different values of $q$. (b) $D_q$ for Werner states,
as a function of the degree of mixedness measured by the linear
entropy, for different values of $q$. All depicted quantities are
dimensionless.\label{Wernerstates}}
\end{center}
\end{figure}

\nd Let us specialize (\ref{ana1}) to the particular
instance $c_1=c_2=c_3=-c$, i.e., the celebrated Werner
states,

\begin{equation}
 \rho=(1-c)\frac{\mathbb{I}}{4}+c|\Psi^-\rangle\langle\Psi^-|,\,\,\,
 c\in[0,1],
\end{equation}

\noindent
with $|\Psi^-\rangle=\frac{1}{\sqrt{2}}(|01\rangle-|10\rangle)$.
By  following \cite{Luo08b} one easily obtains

\begin{eqnarray}
 \vartheta(\rho)=&-&2\left(\frac{1}{2}\right)^q\left[\ln_q\frac{1}{2}+\left(\frac{1-c}{2}\right)^q\ln_q\frac{1-c}{2}+\left(\frac{1+c}{2}\right)^q\ln_q\frac{1+c}{2}\right]\nonumber\\&+&3\left(\frac{1-c}{4}\right)^q\ln_q\frac{1-c}{4}+\left(\frac{1+3c}{4}\right)^q\ln_q\frac{1+3c}{4},
\end{eqnarray}

\noindent
and, as seen in Fig. \ref{Wernerstates}, positivity prevails for the prototype-mixed state.
  In Fig. \ref{Wernerstates} we plot the (normalized) q-discord as a
function of the state-parameter $c$ for different values of $q$
and also as a function of the mixedness-degree as given by the
linear entropy $$S_L=\frac{4}{3}[1-{\text Tr}\rho^2],$$
  trivially related to the purity $\gamma$ of a state via $S_L \, = \, 1 - \gamma.$
 As expected, an inverse
relationship between mixedness-degree and quantum correlations is
displayed. We remark on the single-valuedness of the Werner-relation between q-discord and mixedness, even for $q=1$
%We also plot level surfaces of q-discord for Bell-diagonal states. From these plots we can see for instance how the level surface $D_q=0$ increases increasing the value of $q$. These plots are not shown in this draft.

%\newpage

\subsection{$\alpha$ states }

\nd We also will study the quantum q-discord for the following
one-parameter states

\begin{figure}[h]
\begin{center}
\vspace{0.5cm}
\includegraphics[scale=0.75,angle=0]{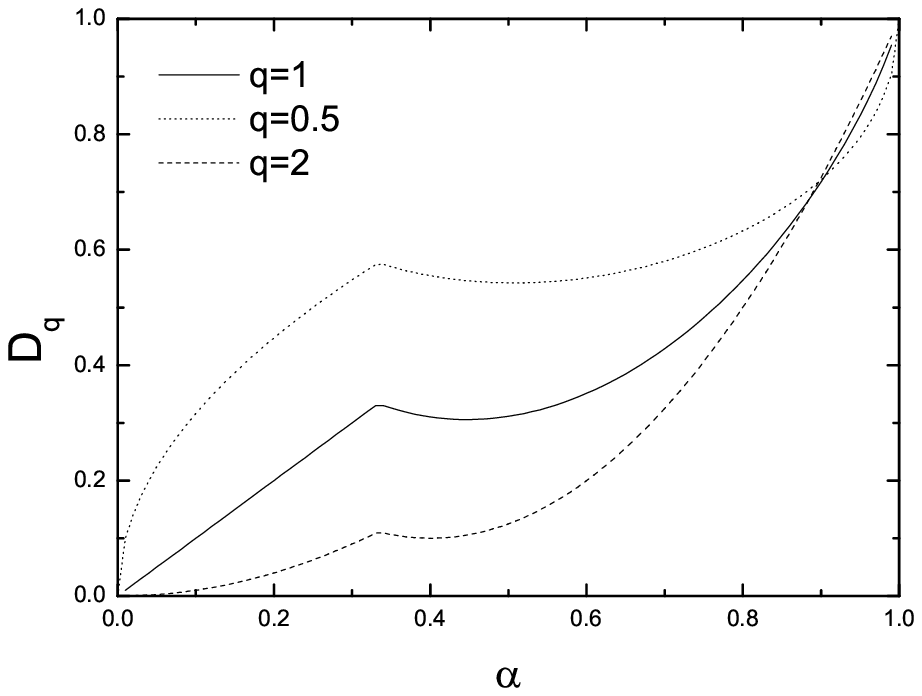}
\includegraphics[scale=0.75,angle=0]{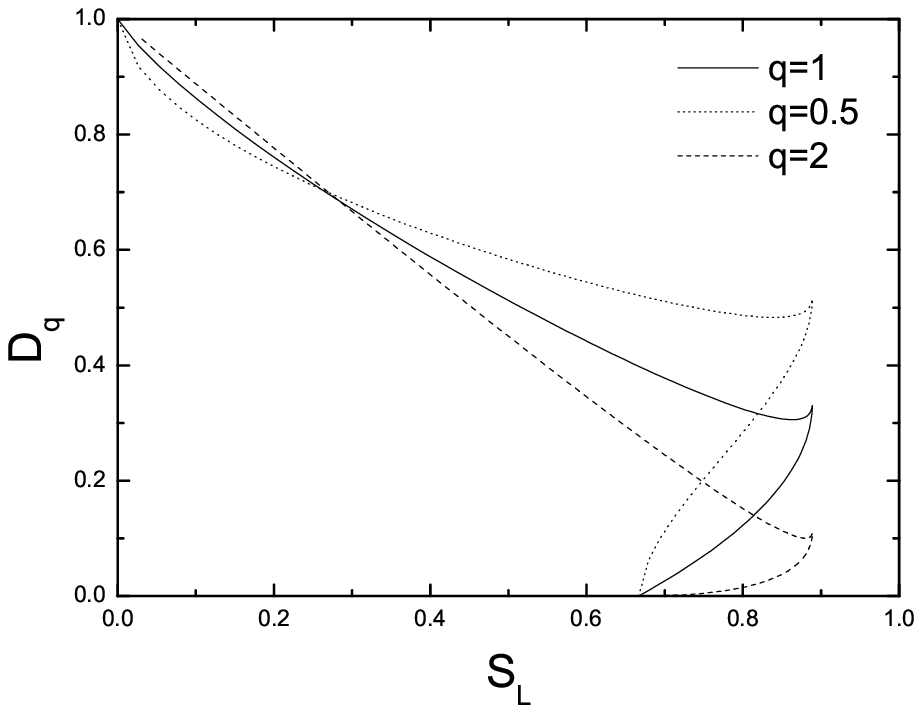}
\vspace{0.5cm} \caption{(a) $D_q$ for $\alpha$ states, as a
function of $\alpha$, for different values of $q$. (b) $D_q$ for
$\alpha$ states, as a function of the degree of mixedness, for
different values of $q$. All depicted quantities are
dimensionless.\label{alphastates}}
\end{center}
\end{figure}

%$\alpha$ and a two parameter ($\alpha,\beta$) states given respectively by:

\begin{equation}\label{alpha-state}
 \rho_{\alpha}=\begin{pmatrix}
  \frac{\alpha}{2} & 0 & 0 & \frac{\alpha}{2} \\
  0 & \frac{1-\alpha}{2} & 0 & 0 \\
 0 & 0 & \frac{1-\alpha}{2} & 0 \\
  \frac{\alpha}{2} & 0 & 0 & \frac{\alpha}{2}
\end{pmatrix}
\end{equation}

\noindent
where $0\leq\alpha\leq 1$. Let $\xi=\max\{|\alpha|,|2\alpha-1|\}$.
The  q-discord becomes

\begin{equation}
 \vartheta(\rho)=-2\left(\frac{1}{2}\right)^q
 \left[\ln_q\frac{1}{2}+\left(\frac{1-\xi}{2}\right)^q\ln_q\frac{1-\xi}{2}+
 \left(\frac{1+\xi}{2}\right)^q\ln_q\frac{1+\xi}{2}\right]+
 2\left(\frac{1-\alpha}{2}\right)^q\ln_q\frac{1-\alpha}{2}+\alpha^q\ln_q\alpha.
\end{equation}
\nd In Fig. \ref{alphastates} we  depict the quantum q-discord as
a function of the state's parameters for different values of $q$
 and also  plot it as a function of the linear entropy.
Positivity again prevails. The single-valuedness between discord and mixedness is lost for these states.

\subsubsection{Discord-differences for two $\alpha-$states}

\begin{figure}[h]
\begin{center}
\vspace{0.5cm}
\includegraphics[scale=0.8,angle=0]{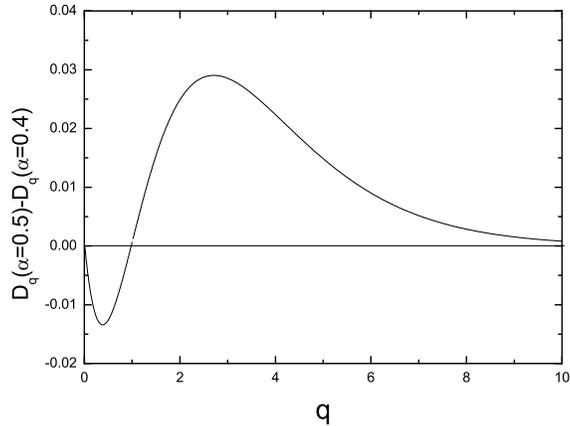}
\vspace{0.5cm} \caption{Difference between $D_q$ of two $\alpha$
states as a function of the parameter $q$. A non trivial ordering
relation is found. All depicted quantities are
dimensionless.\label{alphasigno}}
\end{center}
\end{figure}

\nd In Fig. \ref{alphasigno} we display the difference between the
q-discord of two $\alpha$-states (corresponding to $\alpha=0.4$
and $\alpha=0.5$, respectively),  as a function of $q$. \vskip 3mm
\nd This difference takes negative or positive values depending on
the range of $q$. This is indeed a novel feature. A relation of
order for quantum states based on the discord-concept cannot
univocally be established, because it depends on the entropic
quantifier one chooses to employ. In other words,  the quantal
correlations that the discord quantifies are seen in different
manners by distinct entropic quantifiers. {\it This lack of
uniqueness is the leit-motif of the present considerations}.

%We also study the change in the sign of the derivative function as a function of $\alpha$ for different values of $q$. Different signs in the derivative function indicate an ordering change.

\subsection{$(\alpha,\beta)$ state}

\begin{figure}[h]
\begin{center}
\vspace{0.5cm}
\includegraphics[scale=0.8,angle=0]{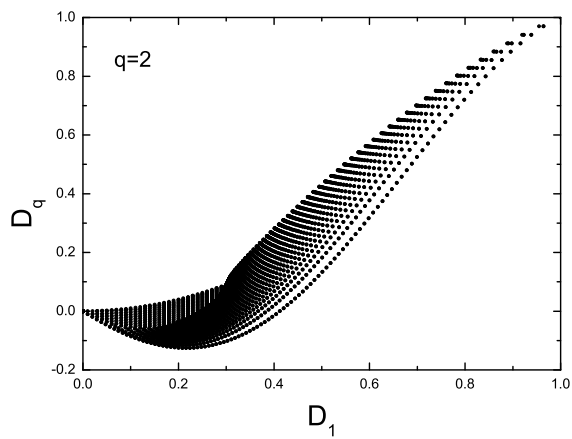}
\vspace{0.5cm} \caption{$D_q$ (q=2) vs $D_1$ for
$\alpha,\beta$-like states. All depicted quantities are
dimensionless.\label{abstate}}
\end{center}
\end{figure}

\nd As a last particular kind of special  state to be analyzed,  consider the two-parameters
state

\begin{equation}\label{alpha-beta-state}
 \rho_{\alpha,\beta}=\frac{1}{2}\begin{pmatrix}
 \alpha & 0 & 0 & \alpha \\
  0 & 1-\alpha-\beta & 0 & 0 \\
 0 & 0 & 1-\alpha+\beta & 0 \\
  \alpha & 0 & 0 & \alpha
\end{pmatrix},
\end{equation}

\noindent
where $0\leq\alpha\leq 1$ and $\alpha-1\leq\beta\leq 1-\alpha$. We
display the $q=2-$discord versus the $q=1-$discord for this state
in Fig. \ref{abstate}. A strong correlation is exhibited between
the two $q-$measures. This could be taken as evidence that
changing $q$ from its original $q=1-$value does not per se modify
the overall manner in which q-discord quantifies quantum
correlations.

\subsection{Arbitrarily mixed two-qubit states}

\nd Here, we focus our discussion on {\sf general} (pure or mixed)
states of two qubits. For such system we can parametrize the
basis of the measurement by $\theta$ and $\phi$,

\begin{eqnarray}
 |\psi\rangle&=&\cos(\theta)|0\rangle+e^{i\phi}\sin(\theta)|1\rangle\\\nonumber
|\psi_{\bot}\rangle&=&e^{-i\phi}\sin(\theta)|0\rangle-\cos(\theta)|1\rangle.
\end{eqnarray}
\begin{figure}[h]
\begin{center}
\vspace{0.5cm}
\includegraphics[scale=0.8,angle=0]{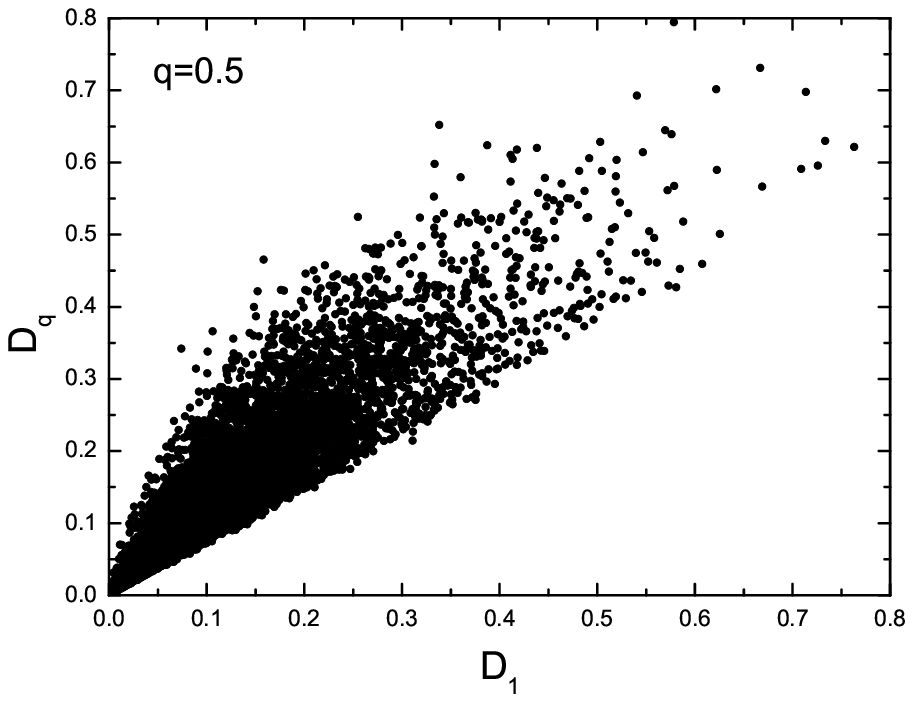}
\includegraphics[scale=0.8,angle=0]{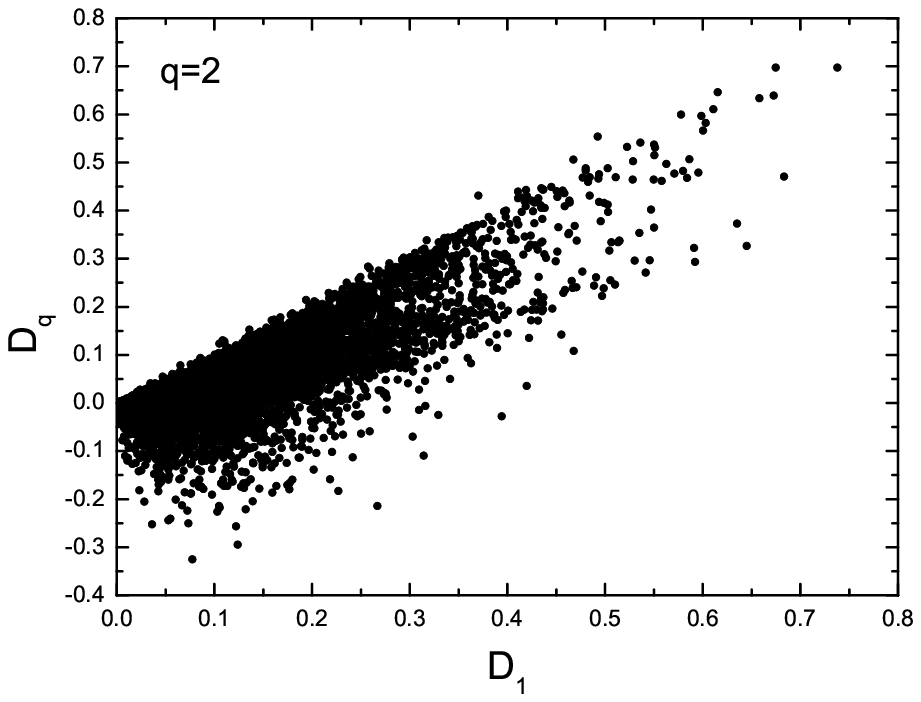}
\vspace{0.5cm} \caption{$D_q$ as a function of $D_1$ for randomly
generated two-qubit states (a) $q=0.5$, (b) $q=2$. All depicted
quantities are dimensionless.\label{qDranstates}}
\end{center}
\end{figure}
\nd We numerically search the $\theta$ - $\phi$ space for the set
of values that maximize Eq. (\ref{optimization}). The resultant
density operator $\rho_{A|\Pi_j^{(B)}}=\rho_j$, when such
measurements are performed on subsystem $B$, is

\begin{equation}
 \rho_j=\frac{1}{p_j}(I\otimes\Pi_j^{(B)})\rho(I\otimes\Pi_j^{(B)}),
\end{equation}

\noindent
where each complete set, composed of two elements,
of possible measurements is defined as follows,

\begin{eqnarray}
 \Pi_1^{(B)}&=&|\psi\rangle\langle\psi|\\\nonumber
\Pi_2^{(B)}&=&|\psi_{\bot}\rangle\langle\psi_{\bot}|.
\end{eqnarray}

\begin{figure}[h]
\begin{center}
\vspace{0.5cm}
\includegraphics[scale=0.8,angle=0]{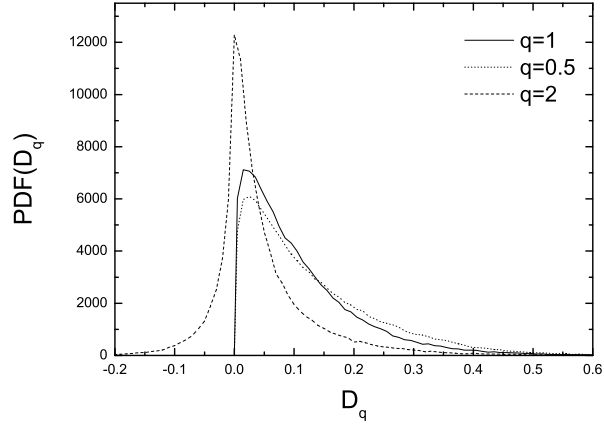}
\vspace{0.5cm} \caption{Probability distribution of finding an
arbitrary two-qubit state with a given value of q-discord for
different values of $q$. The curves were constructed using $10^5$
generated random states. All depicted quantities are
dimensionless. \label{qDPDF}}
\end{center}
\end{figure}
\nd We randomly generate states uniformly distributed according to
the measure $\mu$ and by recourse of the previously described
optimization procedure. Of course, we numerically  search for
$\theta$ and $\phi$ and compute the q-discord for these states. In
Fig. \ref{qDranstates} we display the correlation between the
discord and the q-discord for different values of $q$. Negative
values of the q-discord are depicted in the plot for the case
$q>1$ ($q=2$). Figure \ref{qDPDF} depicts the probability
distribution of finding a given value of q-discord in the whole
space of two-qubit states for $q=0.5,1,2$.

\begin{figure}[h]
\begin{center}
\vspace{0.5cm}
\includegraphics[scale=0.8,angle=0]{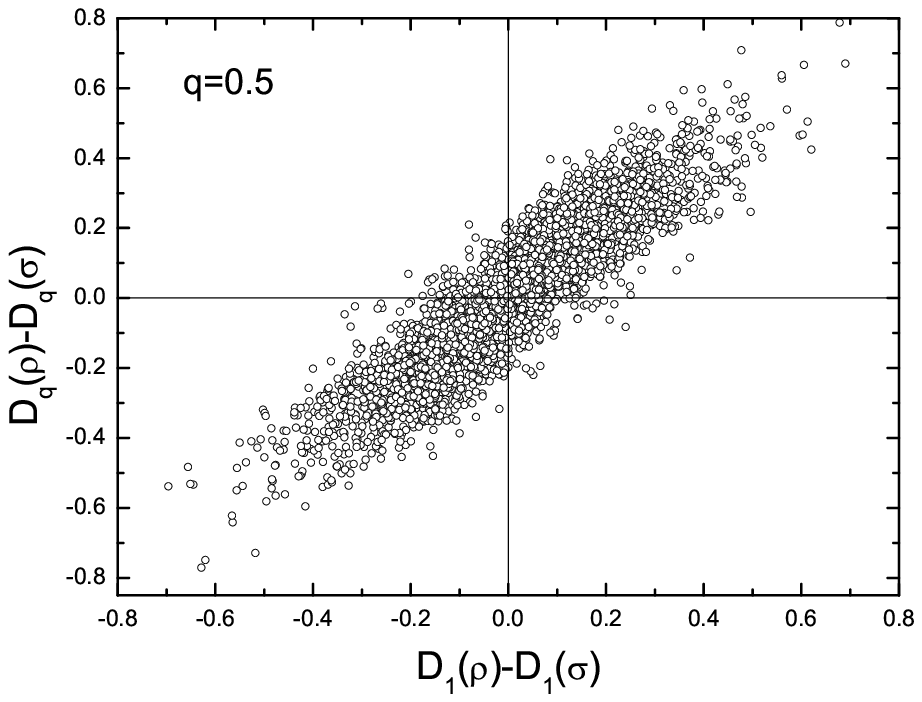}
\includegraphics[scale=0.8,angle=0]{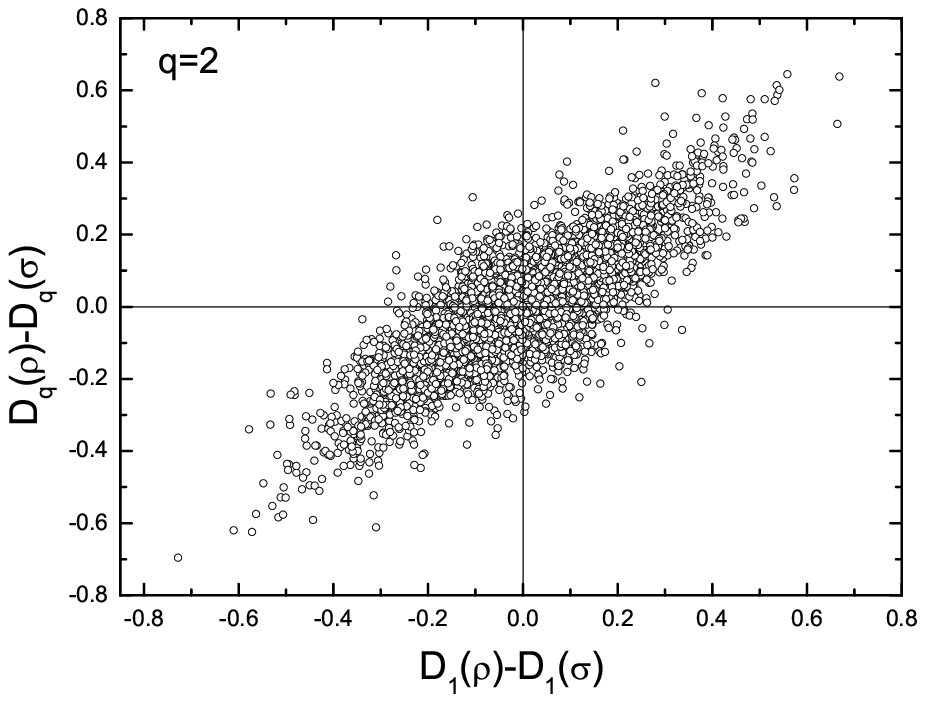}
\vspace{0.5cm} \caption{$D_q(\rho)-D_q(\sigma)$  as a function of
$D_1(\rho)-D_1(\sigma)$ (a) $q=0.5$, (b) $q=2$. All depicted
quantities are dimensionless. \label{qDdif}}
\end{center}
\end{figure}

\nd We also compute the difference between the q-discord and the
discord between pairs of randomly generated states $\rho$ and
$\sigma$. In Fig. \ref{qDdif}
we plot the resultant differences of
the q-discord versus similar differences (for the same pair of
states) corresponding to 1-discord. This plot  depicts the pertinent results.
\vskip 3mm \nd Overall, our numerical simulations confirm the conclusions reached by the analysis of special kinds of states.

\section{Conclusions}

\nd We have introduced a new family of quantum discord
measures that quantifies quantum correlations based in the chain
rule relating the i) conditional- and ii) joint-Tsallis entropies.
  Via two types of study \begin{itemize}
\item of special kinds of quantum states

\item arbitrary, randomly generated mixed states,

\end{itemize}
we have been able to extract the following conclusions:
\begin{enumerate}
 \item There is a strong correlation between the ``new'' q-discord and the original one of Ollivier and Zureck.

\item However, an order-relation for quantum states based on discord lacks unicity because it definitely depends
on the quantifier one chooses to employ. This means that $q$-discord functionals corresponding to different values
of $q$ measure different aspects of the non-classicality (quantumness) of correlations.
\end{enumerate}

\nd
This last fact should constitute strong stimulus for establishing a more detailed assessment
of just what kind of correlations the discord concept quantifies.

\acknowledgments This work was partially supported by the projects
FQM-2445 and FQM-4643 of the Junta de Andalucia and the grant
FIS2011-24540 of the Ministerio de Innovaci\'on y Ciencia (Spain).
A.P.M. acknowledges support by GENIL through YTR-GENIL Program.
A.P. acknowledges support from the Senior Grant CEI Bio-Tic
GENIL-SPR.

\end{document}